\documentclass[aip,jcp,reprint]{revtex4}
\usepackage{amsmath,bm,graphicx}
\usepackage{amsfonts}
\usepackage{graphicx}
\usepackage{epstopdf}
\usepackage{amssymb}
\usepackage{mathrsfs}
\usepackage{color}
\usepackage{latexsym}
\usepackage{hyperref}
\usepackage{wasysym}
\usepackage{dcolumn}
\usepackage{bm}
\usepackage{natbib}
\usepackage{ulem}
\newcommand{\kT}{k_{\text{B}}T}
\newcommand{\Fextv}{{\bf F}_{\text{ext}}}
\newcommand{\Fext}{F_{\text{ext}}}
\newcommand{\gammaeff}{\gamma_{\text{eff}}}

\newlength\figurewidth
\AtBeginDocument{\setlength\figurewidth{0.50\columnwidth}}

\usepackage{color}
\let\omp\marginpar\relax\def\marginpar#1{\omp{\color{red}#1}}

\begin{document}

\title{Active microrheology in corrugated channels}
\date{\today}

\author{Antonio M. Puertas}
\affiliation{Department of Applied Physics, Universidad de Almer\'{\i}a, 04120 Almer\'{\i}a (SPAIN)}

\author{Paolo Malgaretti}
\affiliation{Max-Planck-Institut f\"{u}r Intelligente Systeme, Heisenbergstr. 3, D-70569
Stuttgart, Germany}
\affiliation{IV. Institut f\"ur Theoretische Physik, Universit\"{a}t Stuttgart,
Pfaffenwaldring 57, D-70569 Stuttgart, Germany}

\author{Ignacio Pagonabarraga}
\affiliation{CECAM, Centre Europ\'een de Calcul Atomique et Mol\'eculaire, \'Ecole Polytechnique F\'ed\'erale de Lasuanne, Batochime, Avenue Forel 2, 1015 Lausanne, Switzerland}

\begin{abstract}
We analyze the dynamics of a tracer particle embedded in a bath of hard spheres confined in a channel of varying section. By means of Brownian dynamics simulations we apply a constant force on the tracer particle and discuss the dependence of its mobility on the relative magnitude of the external force with respect to the entropic force induced by the confinement. A simple theoretical one-dimensional model is also derived, where the contribution from particle-particle and particle-wall interactions is taken from simulations with no external force. Our results show that the mobility of the tracer is strongly affected by the confinement. The tracer velocity in the force direction has a maximum close to the neck of the channel, in agreement with the theory for small forces. Upon increasing the external force, the tracer is effectively confined to the central part of the channel and the velocity modulation decreases, what cannot be reproduced by the theory. This deviation marks the regime of validity of linear response. Surprisingly, when the channel section is not constant the effective friction coefficient is \textit{reduced} as compared to the case of a plane channel. The transversal velocity, which cannot be studied with our model, follows the qualitatively the derivative of the channel section, in agreement previous theoretical calculations for the tracer diffusivity in equilibrium. 
\end{abstract}

\maketitle

\section{Introduction}

Understanding transport of ions, molecules, cells and colloids in nano- and micro-fluidic devices is of primary relevance for its biological and technological applications. For example, the transport across synthetic~\cite{Boon2011,Lairez2016,Grebenkov2017} and  biological~\cite{Calero2011,Chinappi2006} channels and pores is controlled by their shape, as well as by the effective interactions between channel walls and the transported objects.
Similarly, in micro- and nano-fluidic circuitry the shape of the channel has been exploited to realize fluidic transistors~\cite{Siwy2008} or diodes~\cite{Guo2013,Balme2017,experton2017ion} and to control ionic~\cite{malgaretti2015,malgaretti2016} and electro-osmotic~\cite{malgaretti2014} fluxes. 

At larger scales, the transport of colloids~\cite{Dagdug2012,Marconi2015,Oshanin2016,Bezrukov2017}, polymers~\cite{Cacciuto2006,Fazli2015,Bianco2016} and even active particles~\cite{Ghosh2013,Dagdug2014,Malgaretti2017} has shown a sensitivity on the geometry of the confining channel. Interestingly, for hard sphere baths, polymer solutions and colloidal suspension the mutual interactions among peers can modulate the ``bare'' transport coefficient provided by the solvent. 
Therefore, in these scenarios the many body effect will play a relevant role in determining the effective transport performance.

Theoretical models have been proposed to describe the dynamics of a tracer particle, both in the active (forced) regime, or in the passive (unforced) case. For the active mode, the problem is typically reduced to one dimension (along the channel axis), the Fick-Jacobs equation, where the channel modulation enters as an effective potential (so-called entropic barrier) \cite{Reguera2006, Burada2008, Burada2010, Bosi2012}. However, a channel with varying section induces anisotropic diffusion, which is only captured when the dynamics is studied in (at least) two dimensions, resulting in a diffusion matrix with non-zero out-of-diagonal terms. In equilibrium\cite{Marconi2015}, it is found that the diffusivity in the longitudinal direction shows a maximum in the channel neck, whereas the transversal diffusion follows qualitatively the derivative of the channel section.

In this work, we study the dynamics of a forced tracer in a colloidal system confined in a corrugated channel with Langevin dynamics simulations and a theoretical model based on the Fick-Jacobs approach. In the simulations, the force has been varied covering the linear and the non-linear regimes while in the theoretical model only the small force regime can be studied. Measuring the steady tracer velocity, allows the determination of the longitudinal and transversal friction coefficients (one diagonal and one out-of-diagonal components of the friction tensor, respectively). Our results show that the effective friction experienced by the tracer particle is strongly affected by both, the geometry of the confining channel and the magnitude of the external force. Surprisingly our data show that effective friction can be \textit{reduced} upon \textit{increasing} the corrugation of the channel, i.e. a plane channel does not provide \textit{optimal} transport. The linear regime at small forces is identified by the linear dependence of the tracer flux with the external force, and allows the application of results from equilibrium. For large forces, the tracer dynamics becomes increasingly dominated by the external force, with a small contribution from the channel corrugation.

\section{Simulation details}

\begin{figure}[h]
\includegraphics[width=\figurewidth]{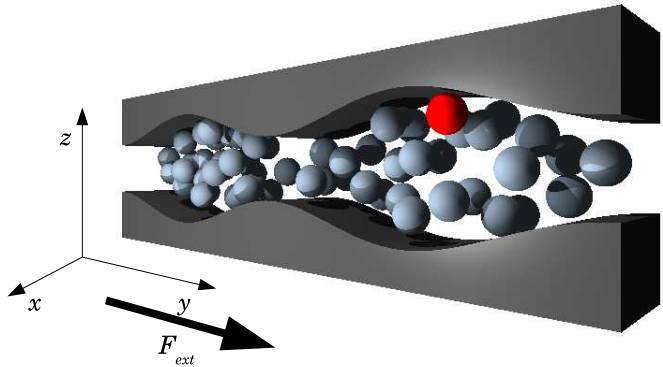}
\caption {Snapshot of the system with $L_z=L_x=6\,a$ and $L_y=40\,a$. The red particle is the tracer. The external force is parallel to the channel mid-plane (y-axis), as shown. \label{snapshot}}
\end{figure}

In the simulations, a system of quasi-hard particles is considered. All particles undergo microscopic Langevin dynamics, which for particle $j$ reads \cite{Dhont1996}:

\begin{equation}
m_j \frac{d^2\, {\bf r}_j}{dt^2}\:=\: \sum_{i\neq j} {\bf F}_{ij} - \gamma_0 \frac{d\, {\bf r}_j}{dt} + {\bf f}_j(t) + {\bf F}_{ext} \delta_{j1}
\label{Langevin}
\end{equation} 

\noindent where $m_j$ is the particle mass, and the terms in the right hand side correspond to {\sl i}) the interaction between particles $i$ and $j$, and with the confining wall, {\sl ii}) the friction force with the solvent, proportional to the particle velocity, {\sl iii}) Brownian force, and {\sl iv}) the external force, which acts only on the tracer (labeled as $j=1$). The Brownian force, ${\bf f}(t)$, is random in time, but its intensity is linked to the friction coefficient, $\gamma_0$, via the Fluctuation Dissipation theorem: $\langle {\bf f}_j(t) \cdot {\bf f}_j(t') \rangle = 6 k_BT \gamma_0 \delta(t-t')$, where $\kT$ is the thermal energy \cite{Dhont1996}. The external force $\Fextv=\Fext \bf{e}_y$ is constant and parallel to the y-axis. The use of Langevin microscopic dynamics to approximate colloidal dynamics is convenient because it gives direct access to the instantaneous velocity ${\bf v}_j=d{\bf r}_j/dt$, which is also used in the equation of motion, and the algorithm is more stable than pure Brownian dynamics due to the inertial term (if time step is smaller than the momentum relaxation time, $m_j/\gamma_0$, collisions are treated as in Newtonian dynamics, while single particle diffusion is obtained at longer times). 

The system is confined in the z-axis between two walls, as shown in the snapshot in Fig. \ref{snapshot}. The shape of the walls is defined by

\begin{equation}
z=\pm \left( \frac{L_z}{2} - A \cos \frac{2\pi y}{\lambda} \right), \label{wall}
\end{equation}

\noindent where $L_z$ is the mean separation between the walls, $A$ is the amplitude of the corrugation and $\lambda$ is its wavelength. The system has periodic boundary conditions in the XY plane (with dimensions $L_x\times L_y$).

The particle-particle and particle-wall interactions are quasi-hard:

\begin{equation}
V(r)=\kT \left(\frac{r}{2a}\right)^{-36} \hspace{0.5cm} \mbox{and} \hspace{0.5cm}
V_w(d)=\kT \left(\frac{d}{a}\right)^{-36}
\end{equation}

\noindent respectively, where $a$ is the particle radius, $r$ is the center-to-center distance and $d$ is the minimal distance from the particle center to the wall. This is calculated expanding the cosine in Eq. \ref{wall} in power series of $2\pi A/\lambda$, up to second order; our simulations are thus valid for small amplitudes and large wavelength. The approximation was validated comparing the density profile of a single particle with the theoretical prediction.

In our simulations, all particles (including the tracer) have the same mass, $m$, and radius, $a$, and the volume fraction of the system is fixed in all cases, $\phi=0.20$. The corrugation amplitude is $A=1\,a$ and its wavelength is $\lambda=20\,a$, and the simulation box  has dimensions $L_x=L_z=6\,a$ and $L_y=2\lambda=40\,a$ (see the snapshot in Fig. \ref{snapshot}). With these parameters, the system contains $N=69$ particles. Despite the low number of particles, there are no significant finite size effects as shown by the results of simulations with $L_x=12\,a$. The length, mass, and energy units are $a=1$, $m=1$,  and $\kT=1$. The solvent friction coefficient is $\gamma_0=10 \sqrt{m \kT}/a$. The equations of motion of the tracer and bath particles were integrated with the Heun algorithm \cite{Paul1995}, with a time step of $\delta t = 0.0005 \, a \sqrt{m/\kT}$. In this algorithm, the friction force is integrated analytically in the time interval $\delta t$.

The tracer is pulled with the constant force $\Fextv$ and dragged through the channel. Due to the periodic boundary conditions, it travels through the simulation box several times. The tracer position and velocity distributions in the $YZ$-plane are recorded in the stationary regime for different forces and analyzed below. Note that the instantaneous velocity, used in the velocity distribution, is well defined since the inertial term is kept in the Langevin microscopic dynamics. With this setup, the non-diagonal component of the diffusion tensor ${\cal D}_{YZ}$ can be determined from the transversal component of the tracer velocity: $v_z = {\cal D}_{YZ} F_{\mbox{ext}}$, in addition to the diagonal component $v_y = {\cal D}_{YY} F_{\mbox{ext}}$.

For comparison purposes, a planar channel has also been simulated with the same particle density and other characteristics; the width of this channel has been set equal to the mean width of the corrugated channel, i.e. $L_z=6\,a$. Some results of this planar channel, as well as for the bulk system with the same density, are given below.

\section{Model}

In order to capture the dynamics of the driven tracer under confinement we extend the Fick-Jacobs approximation to the generic case of  interacting systems. For simplicity's sake we restrict to the case of channels that are translational invariant along the $x$-direction.

The overdamped dynamics of the density of a non interacting system, $\rho(x,y,z,t)$, is described by the Smoluchowski equation:
\begin{equation}
 \dot{\rho}(x,y,z,t)= D \nabla\cdot\left[\nabla \rho(x,y,z,t)+\beta\rho(x,y,z,t)\nabla W(x,y,z)\right]
 \label{eq:smol}
\end{equation}
where $\beta^{-1}=k_B T$ and $W(x,y,z)$ accounts for the geometrical confinement and  for all conservative  forces acting on the particles, as will be described below in detail. When the channel section is varying smoothly, $\partial_y h(y)\ll1$, then the probability distribution can be approximated by~\cite{Zwanzig,Reguera2001}:
\begin{equation}
 \rho(x,y,z,t)=P(y,t)\dfrac{e^{-\beta W(x,y,z)}}{e^{-\beta {\cal F}(y)}}
\end{equation}
\noindent with
\begin{equation}
{\cal F}(y) = -k_BT \ln \left[ \frac{1}{2L_x h_0} \int_{-L_x}^{L_x} dx \int_{-\infty}^{\infty} e^{-\beta W(x,y,z)} dz \right]
\end{equation}
\noindent where $2L_x$ is length of the channel in the $x$ direction (perpendicular to the force) and $h_0$ its mean width. Integrating Eq.(\ref{eq:smol}) in $dx$ and $dz$ leads to 
\begin{equation} 
\partial_t P(y,t)\:=\: \partial_y \left[ \beta D(y) P(y,t) \partial_y {\cal F}(y)+D(y)\partial_y P(y,t) \right] \label{FJ}
\end{equation}
\noindent with $D(y)$ the local tracer diffusion coefficient~\cite{Reguera2001,Marconi2015}.

\paragraph{Non-interacting systems}
In the absence of interactions among the particles, $W$ contains contributions only due to the geometric confinement, interactions with the walls ($\psi$) and the applied external forces ($\Fext$) and can be written as:
\begin{equation}
 W(x,y,z)=
\begin{cases}
\psi(y,z)-\Fext y & \text{if} |z|\leq h(y)\\
\infty & \text{if} |z|> h(y)
\end{cases}
\end{equation}

\paragraph{Interacting systems}
In the mean field approach of the  Fick-Jacobs framework,  particle interactions can be accounted for by terms that are quadratic in the density field.  In this case, 
\begin{equation}
 W(x,y,z)=\begin{cases}
           \int \mathcal{W}(x,z,y,x',y',z')\rho(x',y',z',t)dx'dy'dz'+\psi(x,y,z)-F_\text{ext}y & \sqrt{x^2+z^2}<h(y)\\
           \infty & \text{else}
	  \end{cases}
 \label{eq:W-1}
\end{equation}
can be understood as  the sum of the potential of mean force experienced by the tracer due to the interaction with its peers (via the interaction kernel $\mathcal{W}$, first term in Eq.(\ref{eq:W-1}), the interactions with the walls (second term in Eq.(\ref{eq:W-1}) and the external force (last term in Eq.(\ref{eq:W-1}).

In the steady state , $\partial_t P=0$, Eq. \ref{FJ} becomes a linear first order differential equation, with solution:

\begin{equation}
P(y)=-\frac{J}{D_0} e^{-\beta {\cal F}(y)} \left[ \int_{-L/2}^y \frac{D_0}{D(z)}e^{\beta {\cal F}(z)}dz + \Pi\right] \label{Px-theo}
\end{equation}

\noindent where $D_0$ is the tracer bulk diffusion coefficient. Imposing the normalization of the particle probability density 
\begin{equation}
 \frac{1}{L}\int_{-L/2}^{L/2}P(y)dy=1
 \label{eq:norm}
\end{equation}
and periodic boundary conditions, we can determine $J$ and $\Pi$:

\begin{equation}
J=-D_0 \left[\int\limits_{-L/2}^{L/2}\!\!\! dy e^{-\beta {\cal F}(y)} \left( \int\limits_{-L/2}^y\!\!\! \frac{D_0}{D(z)}e^{\beta {\cal F}(z)} dz + \Pi \right)\right]^{-1}
\label{eq:J}
\end{equation}

\begin{equation}
\Pi=\frac{e^{-\beta {\cal F}(L/2)} \int_{-L/2}^{L/2} \frac{D_0}{D(y)}e^{\beta {\cal F}(y)} dy}{e^{-\beta {\cal F}(-L/2)}-e^{-\beta {\cal F}(L/2)}}
\label{eq:Pi}
\end{equation}

In the following we are interested in the case in which just \textit{one} particle (the tracer) experiences the action of the external force.  Accordingly, the density distribution of the ``passive'' particles is barely affected by the ``active'' motion of the tracer. Hence, for mild values of the external force, the effective potential can be expressed conveniently as
\begin{equation}
 W(x,yz)=W_0(x,y,z)-F_\text{ext}y
\end{equation}
where $W_0$  can be approximated using  the \textit{equilibrium} distribution, leading to
\begin{equation}
{\cal F}(y)= -F_\text{ext}y + {\cal F}_0(y) \label{F}
\end{equation}
with
\begin{equation}
{\cal F}_0(y) = -k_BT \ln \left[ \frac{1}{2L_x h_0}\int\limits_{-L_x}^{L_x} dx \int\limits_{-h(y)}^{h(y)} e^{-\beta W_0(x,y,z)} dz \right]
\label{eq:F0}
\end{equation}
\noindent carrying all the information about the interactions among colloids and with the wall at \textit{equilibrium}. 

Since ${\cal F}_0$ depends on the mutual interactions, it is hard to provide an analytical prediction. In our analysis, ${\cal F}_0(y)$ will be obtained from the simulations, fitting the tracer position distribution from the simulations when no external force is applied. This result for ${\cal F}_0(y)$  is then used for finite forces to calculate both the tracer density and velocity.
Using the proposed  splitting of $\cal F$, $\Pi$ can be rewritten as:

\begin{equation}
\Pi=-\frac{e^{\beta f L/2} \int_{-L/2}^{L/2} \frac{D_0}{D(y)}e^{\beta {\cal F}(y)} dy}{2 \sinh \left( \beta f L/2 \right)} \label{Pi}
\end{equation} 

\noindent which is more useful in the following derivations.

\subsection{Small forces}

In the limit of weak forces, $F_{\mbox{ext}} \ll k_BT/a$, the flux reads
\begin{equation}
J\approx - \frac{D_0}{\Pi} \left[\int\limits_{-L/2}^{L/2} dy \frac{D_0}{D(y)}e^{-\beta {\cal F}(y)}\right]^{-1}\!\!\!\!\!\!\! = \dfrac{\beta D_0 F_\text{ext}}{\int_{-L/2}^{L/2} dy \frac{D_0}{D(y)}e^{\beta \mathcal{F}_0(y)}}
\label{linear-response}
\end{equation}
where we have used the definition of $\Pi$ (Eq.(\ref{eq:Pi})) and the normalization condition, Eq.(\ref{eq:norm}).
Substituting Eq.(\ref{eq:F0}) into Eq.(\ref{linear-response}) and introducing the local equilibrium density profile
\begin{equation}
\rho^\text{eq}_z(y)= \frac{1}{2L_x h_0}\int\limits_{-L_x}^{L_x} dx \int\limits_{-h(y)}^{h(y)} e^{-\beta \psi(x,y,z)} dz
\end{equation}
leads to 
\begin{equation}
 J\approx \dfrac{D_0\beta F_\text{ext}}{\int\limits_{-L/2}^{L/2}\dfrac{D_0}{D(y)}\dfrac{1}{\rho^\text{eq}_z(y)}dy}
 \label{eq:J_expand-1}
\end{equation}

 \begin{widetext}
 
 \begin{figure}[h]
 \includegraphics[width=0.85\figurewidth]{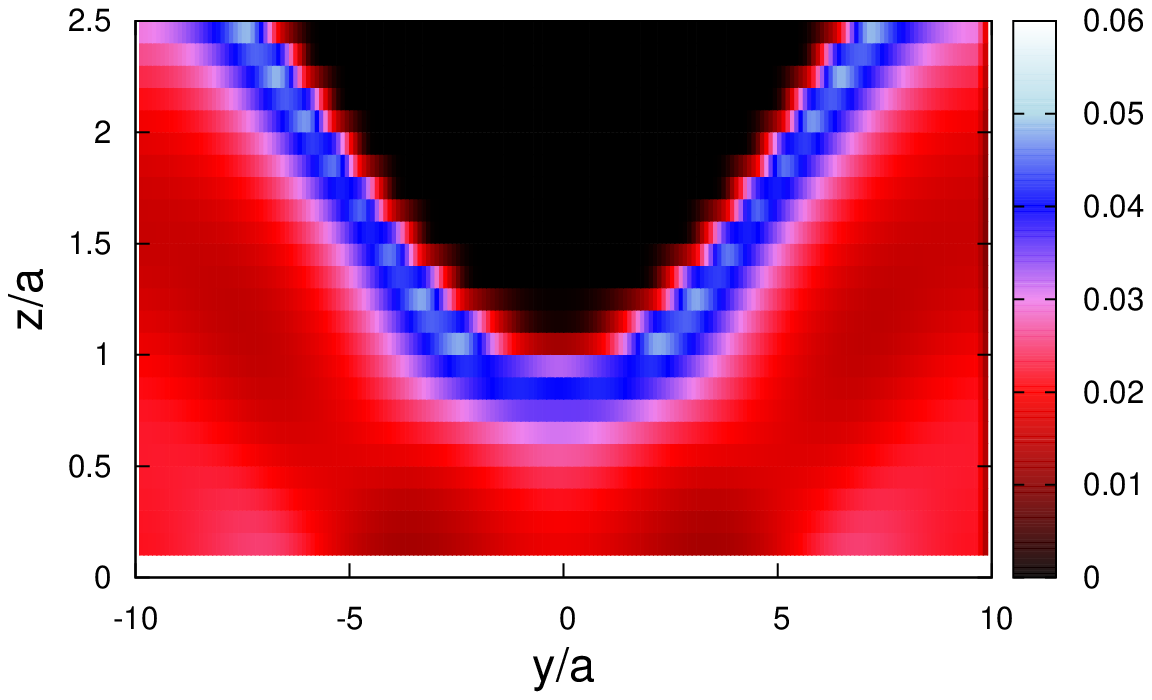}
 \includegraphics[width=0.9\figurewidth]{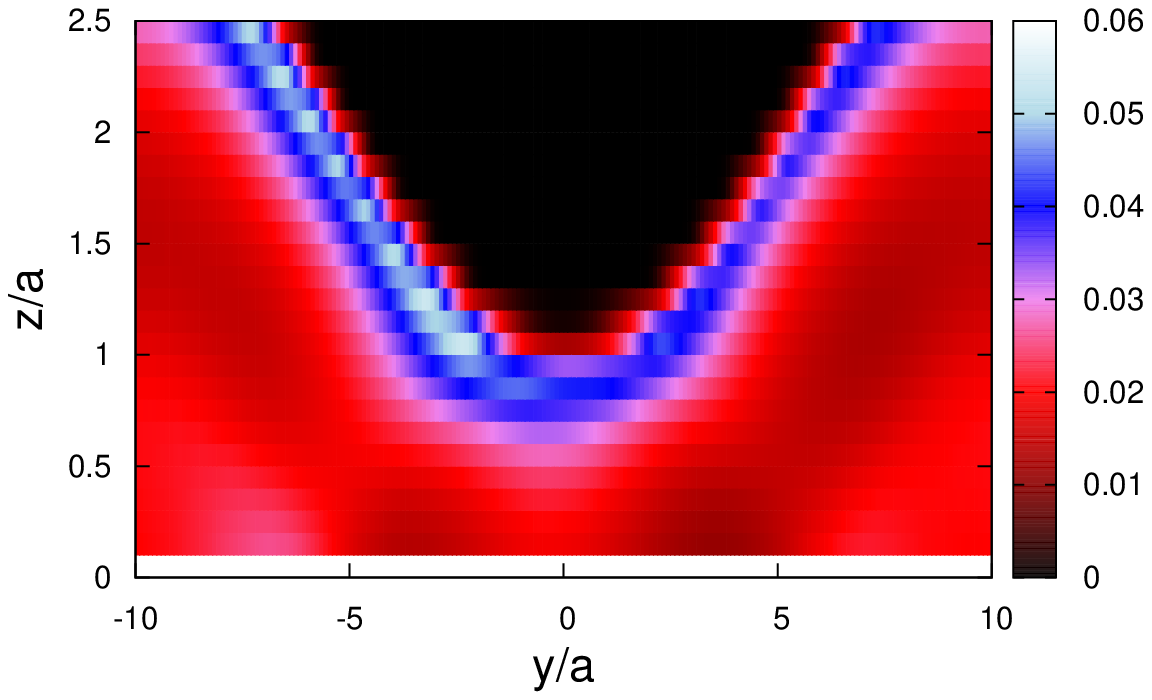}
 \vspace{-0.6cm}
 
 \includegraphics[width=0.9\figurewidth]{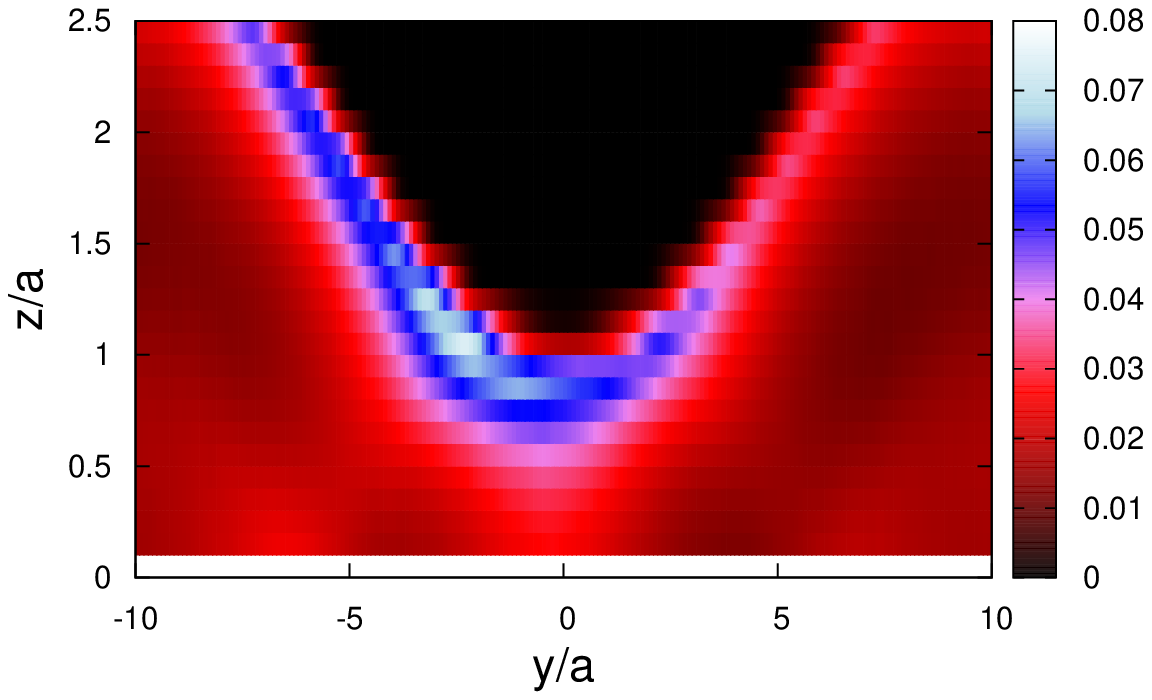}
 \includegraphics[width=0.9\figurewidth]{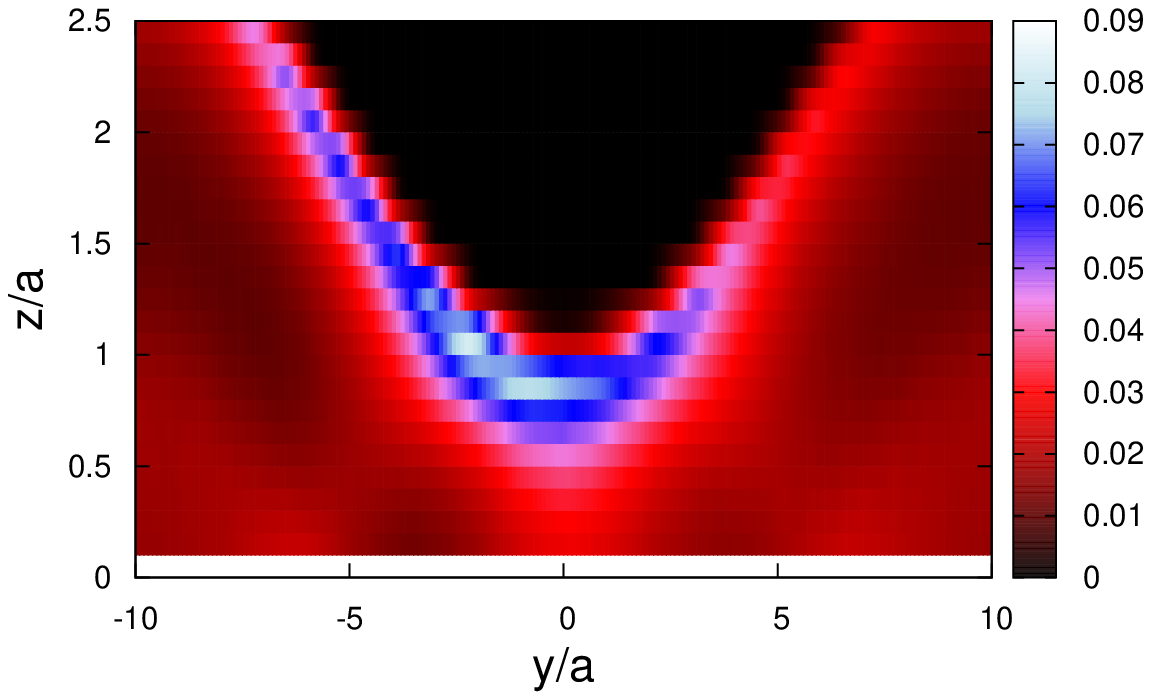}
 \vspace{-0.75cm}
 
 \includegraphics[width=0.9\figurewidth]{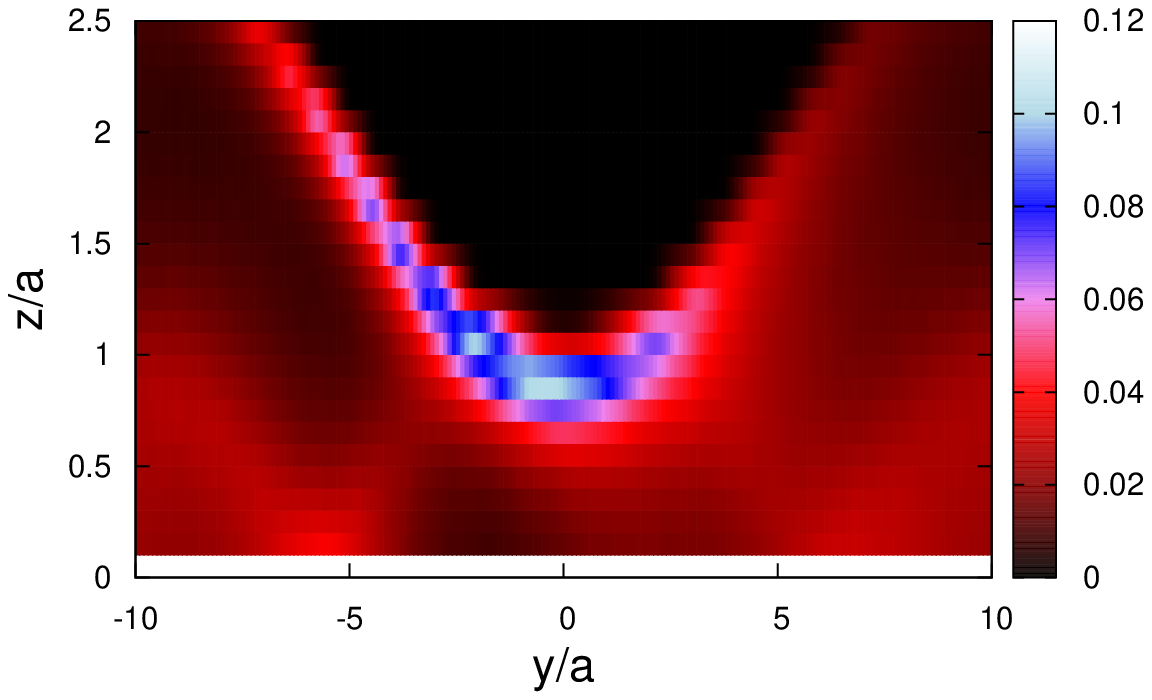}
 \includegraphics[width=0.9\figurewidth]{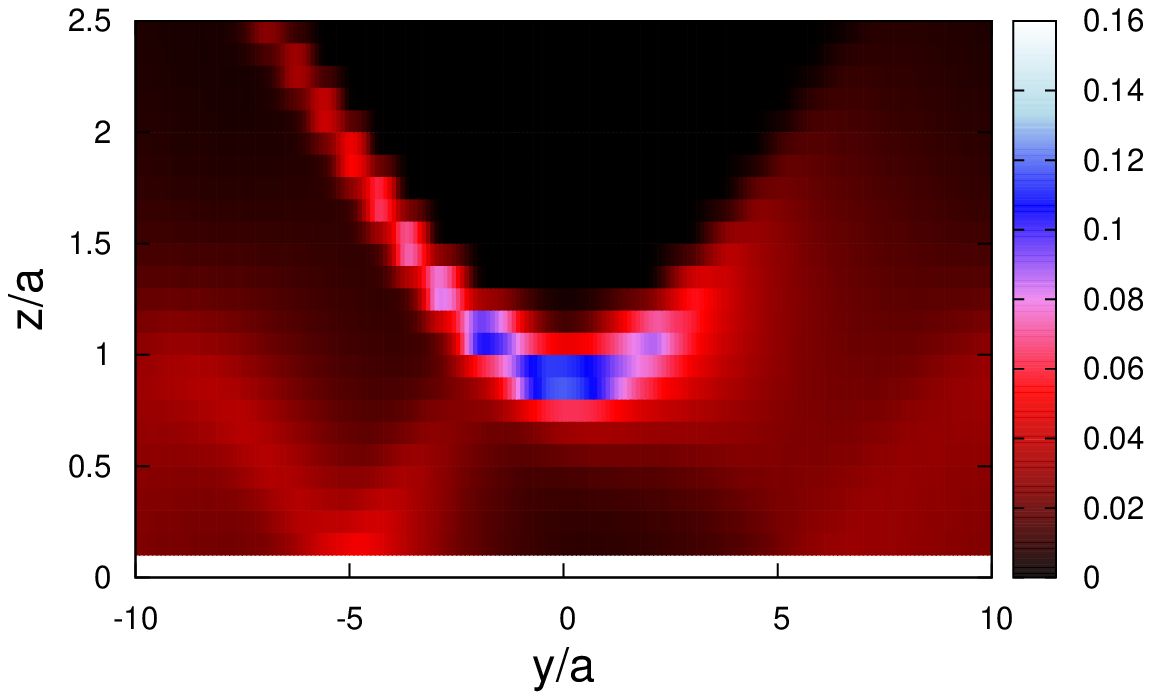}
 \vspace{-0.5cm}
 \caption {Tracer position distributions in the planar channel for different forces, increasing from left to right and top to bottom: $F=0,\,0.1\,\kT/a$ (top row) $0.5,\,2\,\kT/a$ (middle row) and $10,\,50\,\kT/a$ (bottom row). The tracer density increases with the ordering black-red-blue-white.} \label{densityprofile}
 \end{figure}
 \end{widetext}

To proceed, we expand $\rho^\text{eq}_z(y)$ and $D(y)$ around the solution for a plane channel:
\begin{align}
 \rho^\text{eq}_z(y)=\rho_0+\epsilon\rho_1(y)+\mathcal{O}(\epsilon^2)
\end{align}
where $\epsilon$ is a small parameter encoding the deviation of the channel section from the plane case. 
In particular we assume that the modulation of the channel section does not affect the volume of the channel and so the total mass inside the channel is not affected by the modulation. This implies that $\int\limits_{-L/2}^{L/2}dy\rho_1(y)=0$. Similarly we assume that the density, $\varrho$ of the bath of hard spheres can be expanded
\begin{equation}
 \varrho^\text{eq}_z(y)=\varrho_0+\epsilon\varrho_1(y)+\mathcal{O}(\epsilon^2)
\end{equation}
with $\int\limits_{-L/2}^{L/2}dy\varrho_1(y)=0$.
This implies that the diffusion coefficient of the tracer (that is determined by $\varrho$~\cite{Marconi2015}) can be also expanded
\begin{align}
 D(y)=D_0+\epsilon D_1(y)+\mathcal{O}(\epsilon^2)
\end{align} 
Accordingly, to leading order in $\epsilon$ we get
\begin{equation}
 J\approx \frac{\rho_0 D_0\beta F_\text{ext}}{L}   \left[1+ \frac{\epsilon}{L} \int\limits_{-L/2}^{L/2}dy \frac{D_1(y)}{D_0} \right] +\mathcal{O}(\epsilon^2)
 \label{eq:J_expand-2}
\end{equation}
The second factor on the rhs of Eq.(\ref{eq:J_expand-2}) is the modulation in the friction coefficient induced by the geometry. Interestingly, when the diffusion coefficient is independent on the geometry ($D_1=0$), as it is for an ideal gas, Eq.(\ref{eq:J_expand-2}) predicts that for small modulations the effective friction coefficient is independent on the shape of the channel. However, for larger modulations, for which higher order in $\epsilon$ should be considered, Eq.(\ref{eq:J_expand-1}) predicts that the flux across a corrugated channel is \textit{always} smaller than in a flat channel and hence the effective friction is \textit{increased}. In contrast, for interacting systems (for which $D_1\neq 0$), when $\int_{-L/2}^{L/2}\frac{D_1(y)}{D_0}dy>0$ Eq.(\ref{eq:J_expand-2}) predicts that the flux is enhanced as compared to the flat channel and hence the effective friction coefficient is \textit{decreased} by the confinement. 
%


\section{Results}

We start showing the tracer position distribution, and then we move to the analysis of its dynamical properties. In both cases, the simulation results are compared with the theoretical model. Given that the model is expected to fail for large forces, the regimes of small and large forces are presented separately.

\subsection{Tracer position distribution}

The tracer position distributions in the channel, for different values of the external force, are presented in Fig. \ref{densityprofile}. For small and intermediate forces, the tracer distribution mimics the bath equilibrium density. For small forces the maximum probability to find to the tracer is close to the wall (see top right panel), but there is an increasing probability in the bottleneck, and, less noticeable, in the mid-plane of the channel. Eventually, the tracer density strongly deviates from the equilibrium profile for large forces, as shown in the bottom panels of Fig.\ref{densityprofile}. 

\begin{figure}[h]
\includegraphics[width=1\figurewidth]{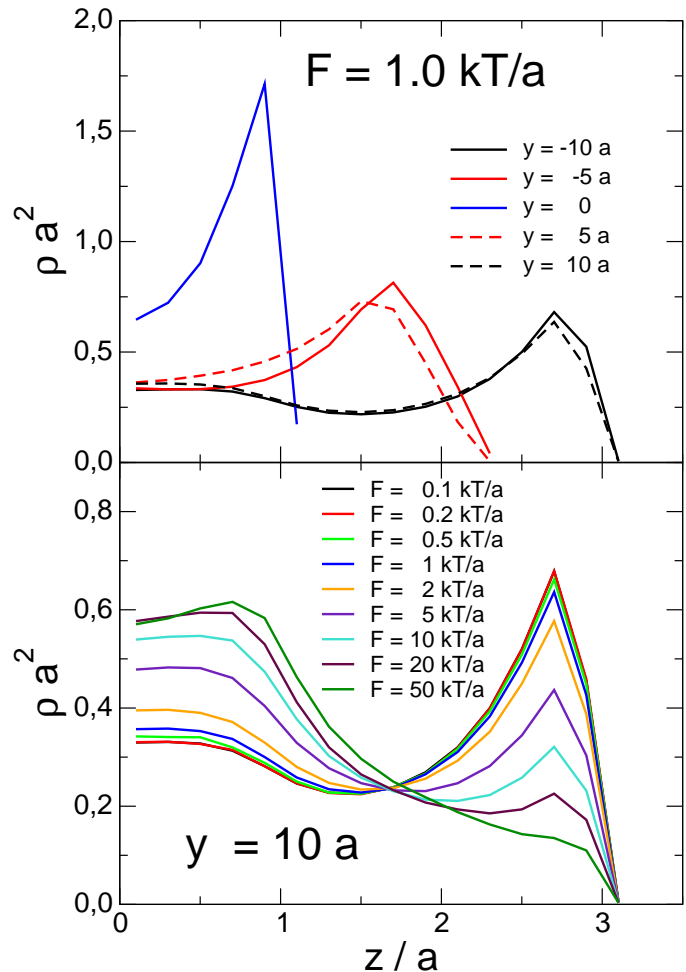}
\caption {Tracer distribution profiles in planes perpendicular to the external force. Different planes are studied in the upper panel, as labeled, for $\Fext=1 \kT/a$, different forces are shown in the lower panel for the widest section ($y=\pm\lambda/2=10\,a$).} 
\label{rhoy}
\end{figure}

In order to analyze quantitatively these distributions, the tracer density is studied in different planes, see Fig. \ref{rhoy}. Since the cross section of the channel varies, the tracer density is studied in slabs of width $\Delta z=1a$. The upper panel of Fig. \ref{rhoy} shows the differences  between the tracer density profiles when the channel narrows or widens, for a mild external force $\Fext a=k_BT$. Similar to the density of bath particles (data not shown) the tracer density is larger close to the walls, with the effect being more pronounced in the neck. 

The lower panel of Fig. \ref{rhoy} shows the tracer position distribution in the widest section of the channel, for different applied forces. As mentioned above, for weak forces, $\Fext a\ll k_BT$, the tracer  accumulates at the walls, similarly to bath particles. However, upon increasing the force, the tracer density decreases close to the wall and increases in the channel mid-plane (around $z=0$). As the velocity of the tracer increases, its translocation time across subsequent bottlenecks becomes smaller than the diffusion time along the transverse direction and the tracer cannot explore the wider parts of the channel, getting effectively \textit{confined} about the mid-plane of the channel.

\begin{figure}
 \includegraphics[scale=0.3]{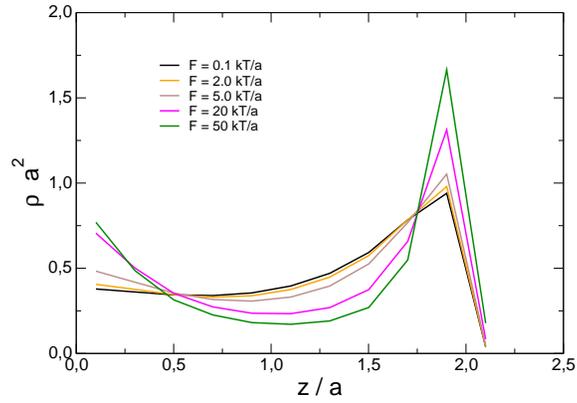}
 \caption{Tracer distribution profiles in a planar channel for different external forces, as labeled.}
 \label{rhoy-planar}
\end{figure}

It is worth studying also an equivalent, uniform and planar channel (Fig. \ref{rhoy-planar}). In this case, 
the tracer position distribution grows close to the wall and in the mid-plane for increasing forces (and decreases for intermediate values of $z$). Namely, in the planar channel, the tracer moves preferentially close to the walls for large forces, contrary to the case of the corrugated channel.

\begin{figure}[h]
\includegraphics[width=1\figurewidth]{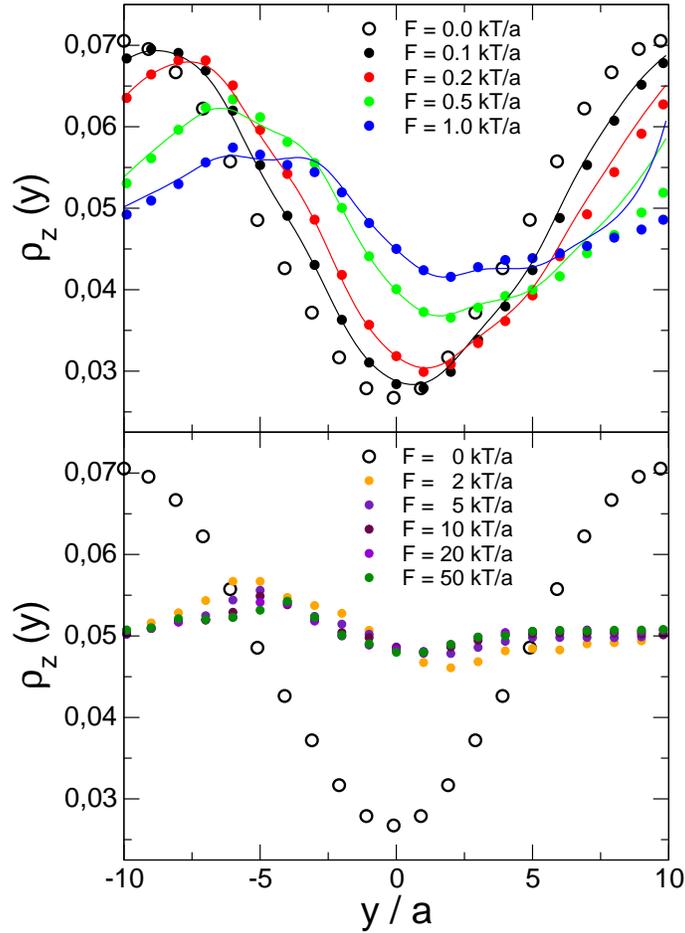}
\caption {Integrated tracer distribution profiles for different forces, as labeled, from simulations (points) and theory (lines). The upper panel shows the low-force regime, and the lower one the large forces. The open circles mark the tracer density for the unforced system.} 
\label{rhoz}
\end{figure}

In order to compare with our model, we focus on the  density profiles integrated in the $XZ$ plane. In the simulations, a slab of width $\Delta y = 1\,a$, parallel to the XZ plane is used to calculate the average:

\begin{equation}
\rho_z(y)= \frac{1}{P_z} \int_{\cal V} \rho(x,y,z) dxdz,
\end{equation}

\noindent where $\cal V$ is the integration volume, and $P_z$ is a constant introduced  to normalize the tracer density, $\int_{-\lambda/2}^{\lambda/2} \rho_z(y) dy=1$. This is presented in Fig. \ref{rhoz} for different values of the external force in the corrugated channel, including the bath density distribution ($F_{\mbox{ext}}=0$). In the latter case, the integrated density is modulated by the channel, with the minimum in the channel neck. For increasing forces, the modulation decreases and displaces to the right, indicating that the channel is explored less efficiently in the transversal direction when the force increases. Interestingly, even the weakest  applied force, $\Fext=0.1\,\kT/a$, provokes noticeable deviations  of the tracer distribution with respect to the  equilibrium bath  density. 

From the equilibrium density profile, ${\cal F}_0$ can be estimated and, once plugged into Eq.(\ref{F}), used to predict the tracer density profile under the action of an external force. The theoretical results, obtained by assuming $D_1=0$, show a good agreement with the simulations for small forces, but deviations are observable for large $y$ above $\Fext=0.5 \kT/a$, due to the accumulation of errors in the numerical integrations implied in the theoretical calculation (see Fig.\ref{rhoz}). For forces above $F_{\mbox{ext}}=1 \kT/a$, the theoretical calculation does no longer predict the tracer behavior.
This underlines the fact that for weak forces the transverse probability distribution is weakly affected by external force and retains its equilibrium form. In contrast for $\Fext>k_BT/a$ the transverse distribution is strongly affected by the external force and the velocity obtained by using the equilibrium transverse distribution does not match the one obtained from simulations.

\subsection{Effective friction with the bath}

\begin{figure}[h]
\includegraphics[width=1\figurewidth]{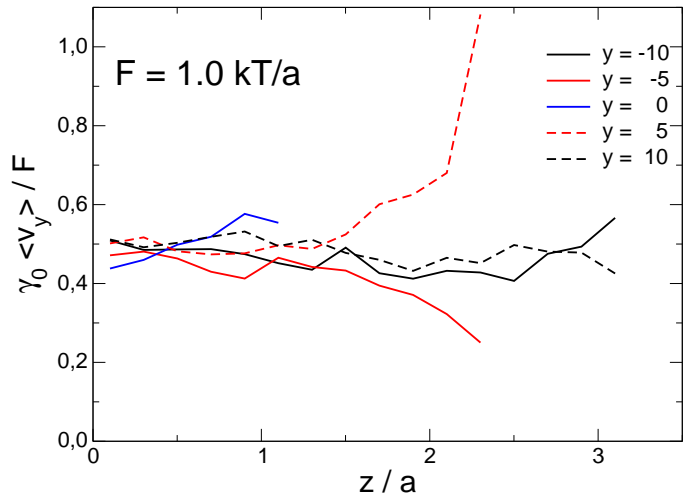}
\caption {Velocity in the force direction, in different planes perpendicular to the external force $\Fext=1\,\kT/a$, as labeled. } \label{vy-XZ}
\end{figure}

We focus now on the tracer dynamics, analyzing its  longitudinal and transversal motion (parallel and perpendicular to the external force, respectively). The distribution of the longitudinal component of the instantaneous tracer velocity in planes perpendicular to the external force is shown in Fig. \ref{vy-XZ}. 

Interestingly, the velocity is constant (within the statistical noise) both in the neck and the widest section, but it varies close to the wall when the channel cross-section is changing. When the channel narrows, the longitudinal velocity is smaller close to the wall, whereas it increases when the channel widens. Additionally, it can be seen in the figure that close to the channel mid-plane, the velocity does not show any dependence on its location  within the channel. In the equivalent planar channel, the velocity is almost constant (increasing slightly in the mid-plane and close to the wall).


\begin{figure}[h]
\includegraphics[width=1\figurewidth]{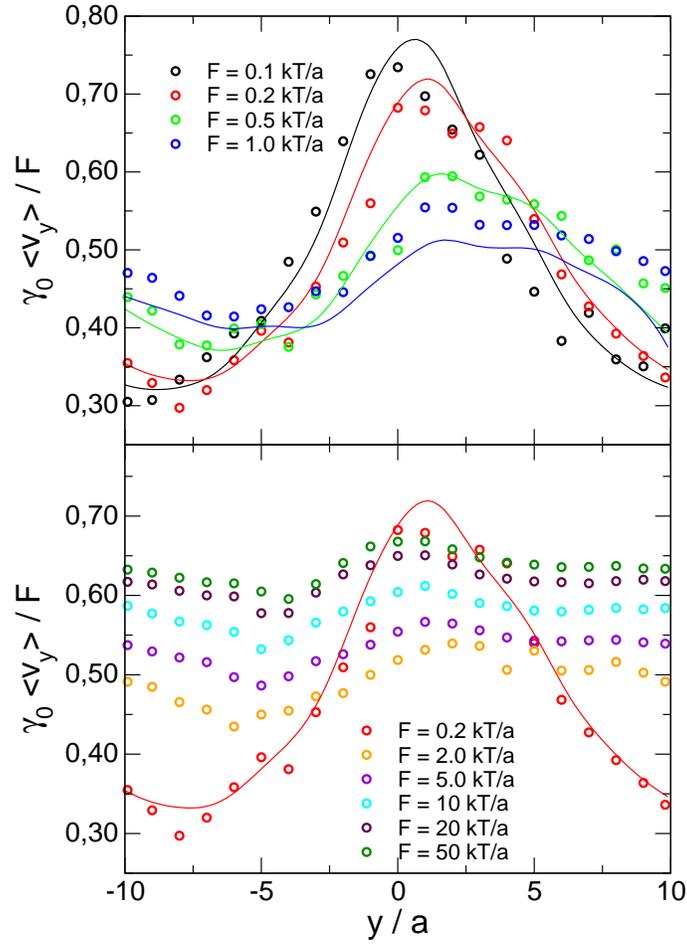}
\caption {Velocity in the force direction, averaged in slabs perpendicular to the channel (symbols show the simulation results, and lines the theory). Small forces are shown in the upper panel, and large forces in the lower one. } \label{vy-allz}
\end{figure}

In order to analyze the impact of the channel constriction on the tracer dynamics, we average its longitudinal velocity in slabs perpendicular to the external force, 
\begin{equation}
\langle v_y (y)\rangle=\frac{1}{\int_{\cal V} \rho(x,y,z) dx\, dz} \: \int_{\cal V} v_y(x,y,z) \rho(x,y,z) dx\,dz
\end{equation}

\begin{figure}[h!]
\includegraphics[width=\figurewidth]{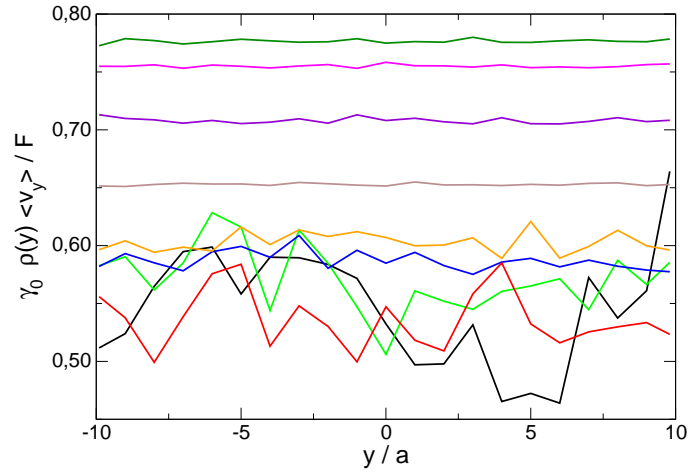}
\caption {Flux in the force direction, with the same color code as the previous figures. Because the flux is divided by the force, the noise is more important for small forces.} \label{flux}
\end{figure}

\noindent Fig. \ref{vy-allz}, presents $\langle v_y \rangle$ from simulations and theoretical predictions as the magnitude of the applied force varies. In order to compare the impact of the channel on the motion of the tracer, we normalize its velocity with the one of an isolated tracer in the bulk, 
 $v_0=F_{\mbox{ext}}/\gamma_0$. The ratio gives us direct information on the tracer longitudinal diffusivity since  $\langle v_y\rangle/v_0=\langle {\cal D}_Y \rangle/D_0 $, where $\langle {\cal D}_Y \rangle$ stands for the average of the local longitudinal tracer diffusivity, ${\cal D}_Y$, over the transverse channel section. The average velocity along the channel for very small forces has a maximum in the narrowest point, but changes notably for increasing force. For large forces, the profile is almost flat, shifting to larger velocities.

The comparison with the theory (thin lines) is possible only for small forces, and the agreement is semi-quantitative. It is worth noting  that the contribution from interactions between particles and with the wall are encoded in ${\cal F}_0$, which is obtained from the tracer density profile in equilibrium. The theory captures nicely both the displacement of the maximum to larger positions, and the decrease of the maximum. For large forces, the contribution of the corrugation can be disregarded and the model predicts a constant velocity profile growing linearly with the force, which is similar to the simulation results. It must be also mentioned that the simulation results are compatible with the theoretical expectation from~\cite{Marconi2015}, where the full diffusion tensor of particles (without external force) in a corrugated channel are calculated. In that case, the diagonal term in the direction of the channel has a maximum in the neck of the channel.

Despite the strong variations of the velocity with the position in the channel, the flux must be homogeneous, as expected in the stationary regime. This is confirmed in Fig. \ref{flux}, where the flux has been calculated as 
\begin{equation}
J(y)=\rho(y) \langle v_y(y) \rangle 
\end{equation}
and normalized with the stationary velocity of the single particle, $F_{\mbox{ext}}/\gamma_0$, in bulk. Within the statistical noise, for small forces the flux is proportional to the force, corresponding to the linear regime shown in Eq. \ref{linear-response}, and increases for large external forces.

\begin{figure}[h]
\includegraphics[bb=1.5 2.0 701.5 489.5,width=\figurewidth]{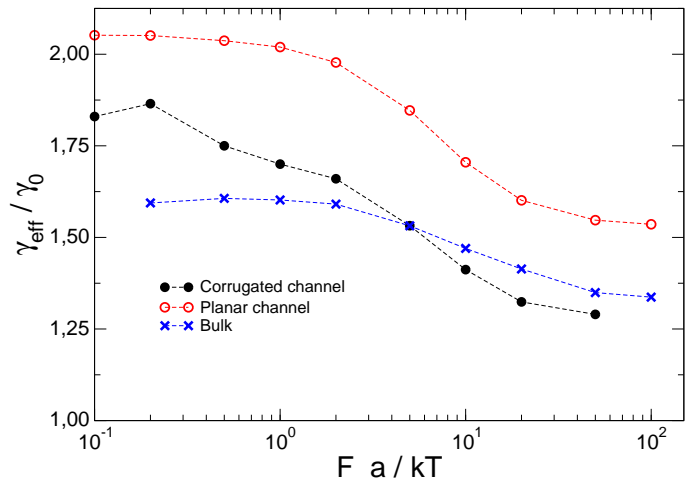}
\caption {Effective friction coefficient in the corrugated channel (black points), compared with the friction observed in the planar channel (red circles) and in the bulk system with the same density and parameters (blue crosses). 
}
\label{gamma}
\end{figure}

The averaged flux provides a robust means to extract the effective friction coefficient experienced by the tracer particle:
\begin{equation}
\gammaeff = \frac{F_{\mbox{ext}}}{\langle J(y) \rangle},
\end{equation}
\noindent Fig. \ref{gamma}  compares the dependence of $\gammaeff$ on the external force for a corrugated channel, an equivalent planar channel and in bulk, the latter for a system with $N=1000$ particles. Overall, the three sets of data follow the same generic trend as that reported for bulk systems \cite{Puertas2014}, with a low-force plateau, force thinning, and (apparently) a high-force plateau.

In the limit of small forces, $\Fext < 1-2 \kT/a$, the friction is almost constant for both channels and for the bulk case, identifying the linear response regime. In this regime the presence of the confining walls induces a larger friction in both channels as compared to the bulk. For large forces, on the other hand, the plateau at high forces is above $1$ in the three sets of simulation data, while the theory sets it at $\gamma/\gamma_0=1$. The origin of this discrepancy was first shown by Squires and Brady in the bulk,  within the theoretical framework of the two-particle Smoluchowski equation \cite{Squires2005}. Comparing the three cases, the planar channel always shows a higher friction coefficient than the corrugated one \footnote{Recall that the channel volume is the same in both cases, i.e. the average section of the corrugated channel equals the constant section of the planar channel.}. Interestingly, a crossover is observed for larger forces, when the friction of the corrugated channel is smaller than that in the bulk system. As shown above, this stems from the effective confinement of the tracer in the central region of the channel, decreasing the friction experienced by the tracer in the corrugated channel. 

Comparing the reduction in the effective friction coefficient shown in Fig.~\ref{gamma} to Eq.~(\ref{eq:J_expand-2}) (and assuming linear response) pinpoints that the contribution of $D_1$ is crucial in determining the flow, even in the regime of very small forces. Indeed, Eq.~(\ref{eq:J_expand-2}) shows that in order to reduce the friction coefficient by means of corrugating the channel, $D_1\neq 0$ is a necessary condition. However, in the same regime, Fig.~\ref{vy-allz} shows that the theoretical model provides quantitatively reliable predictions even assuming $D_1=0$. Hence our data show that the different observables can display quite a different sensitivity to $D_1$.

\begin{figure}[h]
\includegraphics[width=1\figurewidth]{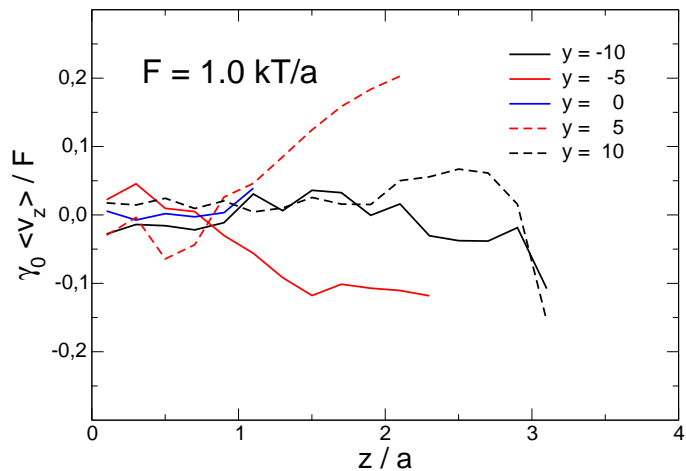}
\caption {Velocity in the transversal direction, in different planes perpendicular to the external force $\Fext=1\,\kT/a$, as labelled. } \label{vz-XZ}
\end{figure}

Finally, we analyze the transversal component of the velocity, giving the non-diagonal component of the diffusion tensor, ${\cal D}_{YZ}$. In the planar channel this component vanishes identically (not shown), while this is not the case for the corrugated channel, as shown in Fig. \ref{vz-XZ}.

The depicted  transversal velocity, normalized by the corresponding tracer velocity in the bulk,  deviates from zero more significantly close to the walls when the channel widens or narrows (following the wall direction). Close to the mid-plane, and both to the neck  and the maximal aperture positions, this velocity component vanishes. This result cannot be discussed within the simple theory model used above, and we must turn to the full model presented previously \cite{Marconi2015}. The theoretical results (for the unforced tracer) indeed predict this behavior close to the wall.

\begin{figure}[h]
\includegraphics[width=\figurewidth]{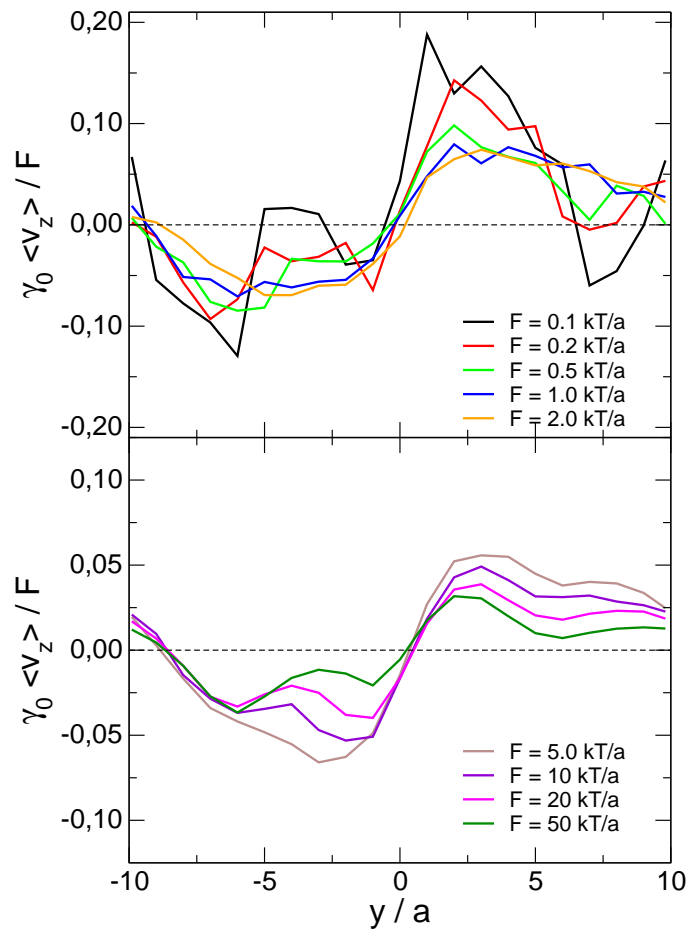}
\caption {Velocity perpendicular to the force direction, averaged in slabs parallel to the XZ-plane. Small forces are shown in the upper panel, and large forces in the lower one. Note that in the averaging, the sign of $v_z$ has been corrected; positive represents from the mid-plane to the wall, and vice-versa.}\label{vz-allz}
\end{figure}

Fig. \ref{vz-allz} presents the transversal velocity averaged in slabs perpendicular to the external force, as studied above for the density and longitudinal component of the velocity. (To avoid a vanishing average due to the channel symmetry, $v_z$ has been defined as $v_z={\bf v}\cdot \hat{\bf n}$, where $\hat{\bf n}$ is the unit vector in the vertical direction pointing from the mid-plane to every point -- upwards in the upper channel half, and downwards in the lower one--). Again, to compare different forces, the velocity is normalized with the  longitudinal velocity of an isolated tracer (i.e. the ratio of the transversal diffusivity to the diffusion constant of the single particle). For small forces (upper panel), in the linear regime, the results collapse onto a master curve, which follows, qualitatively, the derivative of the wall. For large forces, on the other hand, the behaviour changes; the effect disappears with increasing forces, indicating that the tracer motion is increasingly confined in the transversal direction and confirming that the tracer does not explore the full section of the channel.

\section{Conclusions}

The dynamics of a tracer pulled through a colloidal system confined in a corrugated channel has been analyzed. The tracer is pulled with a constant force, and the whole range of forces, has been studied. The results are compared with a simple one dimensional model based on the Fick-Jacobs approximation, but the results from the full model, studied previously, have also been considered; the latter predicts that for the unforced tracer the diffusion tensor has non-zero out-of-diagonal terms.

Our simulations confirm these predictions in the limit of small forces, and show that the linear response regime extends up to $\Fext \sim 1-2\,\kT/a$. The tracer longitudinal velocity has a maximum in the neck of the channel, whereas the transversal component is non-zero and has a maximum where the channel cross-section varies more strongly. Likewise, in this region both the longitudinal and transversal velocities (or local diffusion constants) vary close to the walls, while they remain essentially constant in the rest of the channel.

The theoretical model describes the tracer dynamics effectively, fitting the contribution from particle-particle and particle-wall collisions in the equilibrium case ($\Fextv=0$), and using this result for finite forces. The results for the tracer density and longitudinal diffusivity agree almost quantitatively with the simulation results within the linear regime. Outside this, the calculations break down and cannot provide reliable results.

For larger forces, the tracer is confined to a narrow region parallel to the channel axis, set by the minimum cross-section of the channel. The longitudinal component of the velocity in this region is  almost constant, as the channel section is not explored, and the transversal component becomes increasingly small. As a result, the effective friction experienced by a tracer pulled with a large force in the corrugated channel is smaller than in the bulk with the same density. This region falls out of the theoretical model developed here.


\acknowledgments

This work has been supported by the Spanish Science and Technology Commission (CICYT), the European Regional Development Fund (ERDF) under contracts FIS2015-69022-P and FIS2015-67837-P and by the COST Action MP1305 “Flowing Matter”. I.P. acknowledges support from DURSI Project 2017SGR-884, and SNF project 200021-175719


\bibliography{biblio}

\end{document}